# Role of rare-earth ionic radii on the spin-phonon coupling in multiferroic ordered double perovskites

R. B. Macedo Filho[1,2], D. A. B. Barbosa[3], H. Reichlova[4,5], X. Marti[4,5], A. Menezes[3], A. P. Ayala[2] and C. W. A. Paschoal[3,*]

[1] Instituto Federal de Educação, Ciência e Tecnologia do Maranhão, IFMA, 65030-005 São Luıs - MA, Brazil.

[2] Departamento de Física, Universidade Federal do Ceará, Campus do Pici, PO Box 6030, 60455-970, Fortaleza - CE, Brazil.

[3] Departamento de Física, Universidade Federal do Maranhão, Campus do Bacanga, 65085-580, São Luis- MA, Brazil.

[4] Institute of Physics ASCR, v.v.i., Cukrovarnická 10, 162 53 Praha 6, Czech Republic

[5] Centre d'Investigacions en Nanociencia I Nanotechnologia (CIN2), CSIC-ICN, Bellaterra 08193, Barcelona, Spain

## Abstract

In this paper we investigated the influence of the rare-earth ionic radii on the spin-phonon coupling in $RE_2NiMnO_6$ double perovskites by Raman spectroscopy. Spin-phonon in dense $Nd_2NiMnO_6$ and $Gd_2NiMnO_6$ ceramics were investigated by Raman spectroscopy at low temperatures. The magnitude of the coupling observed in comparison with other isostructural compounds shows that it is not influenced by the rare-earth ionic radius, as well as the deviation of the position of the stretching phonon in the ferromagnetic phase with relation to the anharmonic contributions follows a power law.

[*] Corresponding Author: C. W. A Paschoal: e-mail: paschoal.william@gmail.com; Tel +55 98 3301 8291

Rare-earth based manganites with double perovskite structure has been intensively investigated in last decade since the discovery of magnetocapacitance in La$_2$NiMnO$_6$ (LNMO) by Rogado et al [1]. This compound is a multiferroic, whose Curie temperature is near room temperature ($T_C \sim 280$ K), which permits its application in devices whose magnetic properties can be controlled by electric field.

Since them, an intense attention has been given to electric and magnetic properties of LNMO in several kinds of samples, as ceramics, single crystals, nano-powder and thin films [2–8]. In addition, it was observed a huge spin-phonon coupling in thin films of LNMO [9,10], which it is an important feature for magnetoelastic applications. However, the investigations about spin-phonon coupling were mostly focused in LNMO, while important questions keep open for other compounds of the family, as if the magnitude of the spin-phonon coupling depends on the rare-earth ionic radii.

In addition to La$_2$NiMnO$_6$, spin-phonon coupling was already observed in Pr$_2$NiMnO$_6$ (PNMO) [11], Tb$_2$NiMnO$_6$ (TNMO) [12] and Y$_2$NiMnO$_6$ (YNMO) [13], all in ceramic samples. In the TNMO case, the authors observed a weak spin-phonon coupling, which could not be observed in the stretching phonon position, as it is usually observed in double perovskite, but just in its linewidth [12]. They attributed the low effect to the increasing of rare earth ionic radii. However, the effect was stronger in YNMO [13], in which the effect was similar to that observed in PNMO. That result suggested the ionic radii not strongly influence the spin-phonon coupling magnitude. Recently, spin-phonon coupling observations in Gd(Co,Mn)O$_3$ suggested the B-site order in double perovskite is the main parameter influencing the spin-phonon coupling [14], as expected once this kind of disorder increases the antiferromagnetic interactions through a superexchange mechanism. In this letter, we investigated the ceramics $Nd_2NiMnO_6$ (NNMO) and $Gd_2NiMnO_6$ (GNMO), synthesized at similar conditions of YNMO[13], in order to determine the ionic radii influence on the magnitude of spin-phonon coupling in rare-earth based manganites with double perovskite structure.

Polycrystalline samples of NNMO and GNMO were synthesized by a solid state route according to a stoichiometric mixture of $RE_2O_3$, NiO, and MnO oxides. The samples were calcined ten times at 1000°C for 12 h. The samples were grounded in an agatha mortar and sieved to obtain a homogeneous powder between each calcination. Finally, the samples were sintered at 1400 °C for 48 h. The crystalline structures were probed by X-ray powder diffraction using a Bruker diffractometer model D8 Advance, in a continuous scanning mode using Cu-Ka radiation (40 kV, 40 mA, 0.02 $\theta s^{-1}$). The phases were refined using the Fullprof suite [15]. The refined X-ray powder diffraction patterns obtained for NNMO e GNMO showed that both compounds exhibit an ordered monoclinic-distorted double perovskite lattice, which belongs to the $P2_1/n$ space group, as it is shown in Figure S1 of the supplementary material. The refined structure parameters are in excellent agreement with those obtained previously by Yang *et al*[16], as well as with those predicted by SPuDS code [17]. The refinement and structural parameters are summarized in Table SI (see supplementary material). Virtually unavoidable traces of starter oxides were detected as secondary phase, which are routinely observed in the synthesis of rare-earth mixed perovskites [12,13], and are indicated as * in Figure S1. It is important to point out the Ni-O-Mn angle in both compounds favors the superexchange interaction.

Magnetic measurements were carried out in Quantum Design (QD) superconducting quantum interference device (SQUID). Temperature sweeps were collected with 4 cm long Reciprocating Sample Option scans. The temperature-dependent magnetization and susceptibility reciprocal of NNMO and GNMO (see Figure S2 of the supplementary material) clearly show that both samples exhibit ferromagnetism around $T_C$ = 150 K and 210 K, respectively, which agree with the result obtained by Booth et al [18], and a low coercive field. Low coercive fields usually imply in double perovskites with high B-site ordering. Also, the absence of a second magnetic transition below the observed $T_C$ indicates the order is NNMO and GNMO is high, as happened for YNMO prepared under the same conditions[13].

Raman spectroscopy measurements were performed using a Jobin-Yvon T64000 Triple Spectrometer configured in a backscattering geometry coupled to an Olympus Microscope model BX41 with a 20x achromatic lens. An Innova Coherent laser operating at 100 mW emitting in a 514.5nm line was used to excite the Raman scattered signal, which was collected in a $N_2$-cooled CCD detector. All slits were set up to achieve a resolution lower than 1 cm$^{-1}$. Low-temperature measurements were performed by using a closed-cycle He cryostat where the temperature was controlled to within 0.1 K.

Figures 1a and 1b show the low-temperature Raman spectrum of NNMO and GNMO at 40 K. We observed 8 and 4 modes for NNMO and GNMO, respectively. For this monoclinic structure, there are 24 Raman active modes, which can be described with basis on the oxygen octahedra internal modes as [19]: $6T(3A_g \oplus 3B_g) + 6L(3A_g \oplus 3B_g) + 2\nu_1(A_g \oplus B_g) + 4\nu_2(2A_g \oplus 2B_g) + 6\nu_5(3A_g \oplus 3B_g)$, where $L$ and $T$ are lattice modes (translational, T, and librational, L), while $\nu_1$, $\nu_2$ and $\nu_5$ are oxygen octahedra internal modes, which are expected to be observed for wavenumber higher than 380 cm$^{-1}$. The observed modes can be assigned with basis on the normal modes calculations for

La$_2$CoMnO$_6$ [20] and Gd(Co,Mn)O$_3$ [14], as well as on the other double perovskite manganites previously measured [13]. Thus, for NNMO, the symmetric ($v_1$) and antisymmetric ($v_2$) were observed at 658 and 449 cm$^{-1}$ respectively, while the bending was observed at 521 cm$^{-1}$. In GNMO case, the symmetric ($v_1$) and antisymmetric ($v_2$) were observed at 650 and 465 cm$^{-1}$ respectively, while the bending was observed at 518 cm$^{-1}$. However, it is important point out the significant difference between the number of modes observed for NNMO and GNMO. As reported in YNMO case, GNMO (eight modes observed) shows a higher number of modes compared to NNMO (just four modes observed). This result confirms the higher lattice monoclinic distortion for compounds with rare earths with low ionic radii. GNMO and YNMO (observed 13 Raman-active modes)[13] show a clear splitting of the modes that originated from cubic aristotipic symmetry. A summary of the observed Raman-active phonons is given in Table SII of the supplementary material.

The temperature-dependent Raman spectra of NNMO and GNMO are depicted in Figure 1(c) and 1(d). As it is usual in double perovskite manganites that do not undergo structural phase transitions no anomalies were observed in the temperature range investigated[12-14,21,22]. The temperature dependence of the stretching phonon positions is shown in Figure 2 for NNMO and GNMO. Usually, under temperature changes, the position of phonon follows the Balkanski's model, which consider the anharmonicity contributions according to the expression

$$\omega(T) = \omega_o - C\left[1 + \frac{2}{(e^{\hbar\omega_o/k_BT}-1)}\right] \qquad (1)$$

with $C$ and $\omega_o$ being fitting parameters. When considered $\nu_1$ stretching mode, the stretching, the behavior departs considerably of the Balkanski's model below the Curie temperature of the samples, as shown in Figure 2.

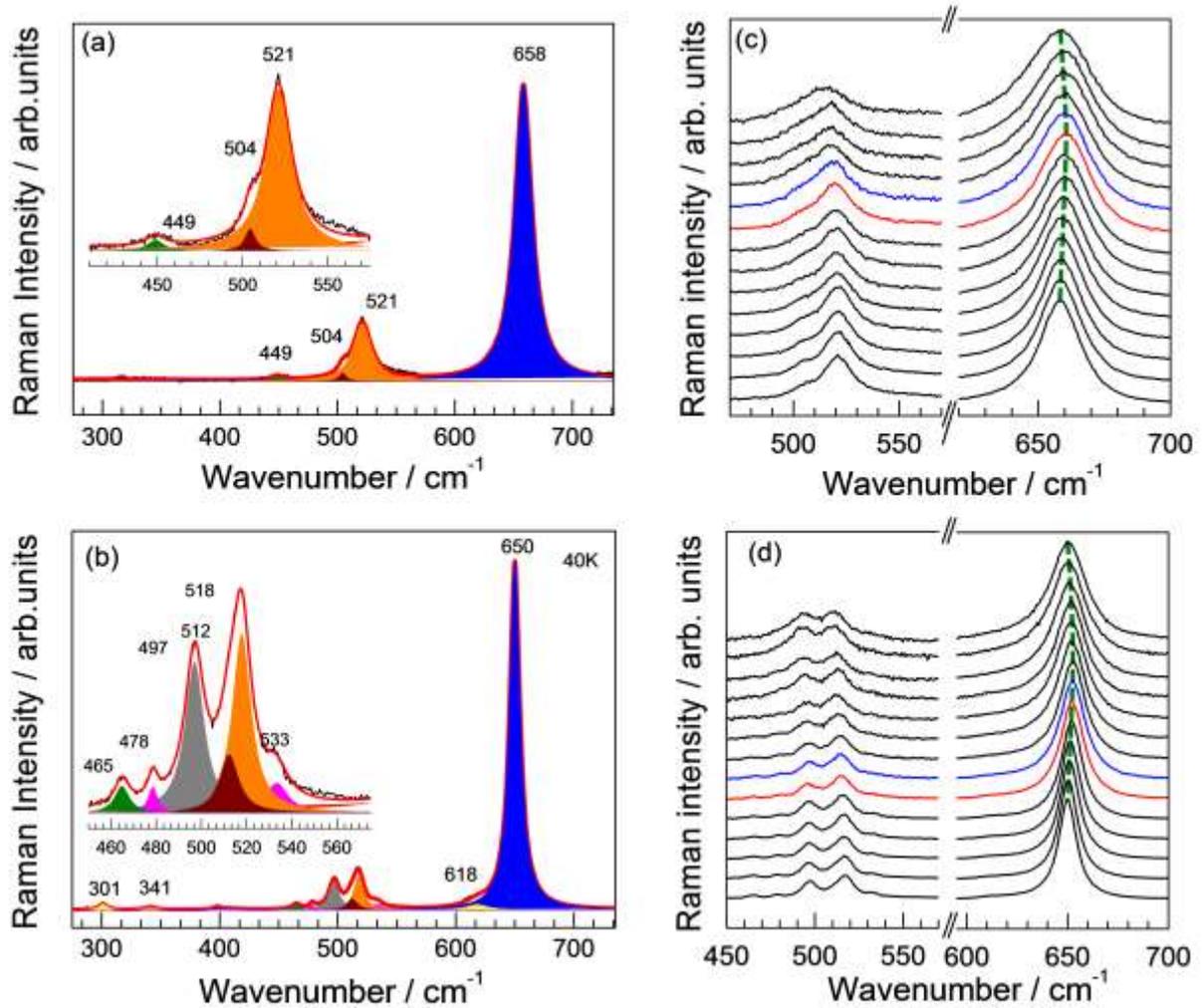

Figure 1 – Low-temperature Raman spectrum of NNMO (a) and GNMO (b) at 40 K. Temperature dependence of Raman spectra of NNMO (c) and GNMO (d) from 40 K up to 300 K in steps of 20 K.

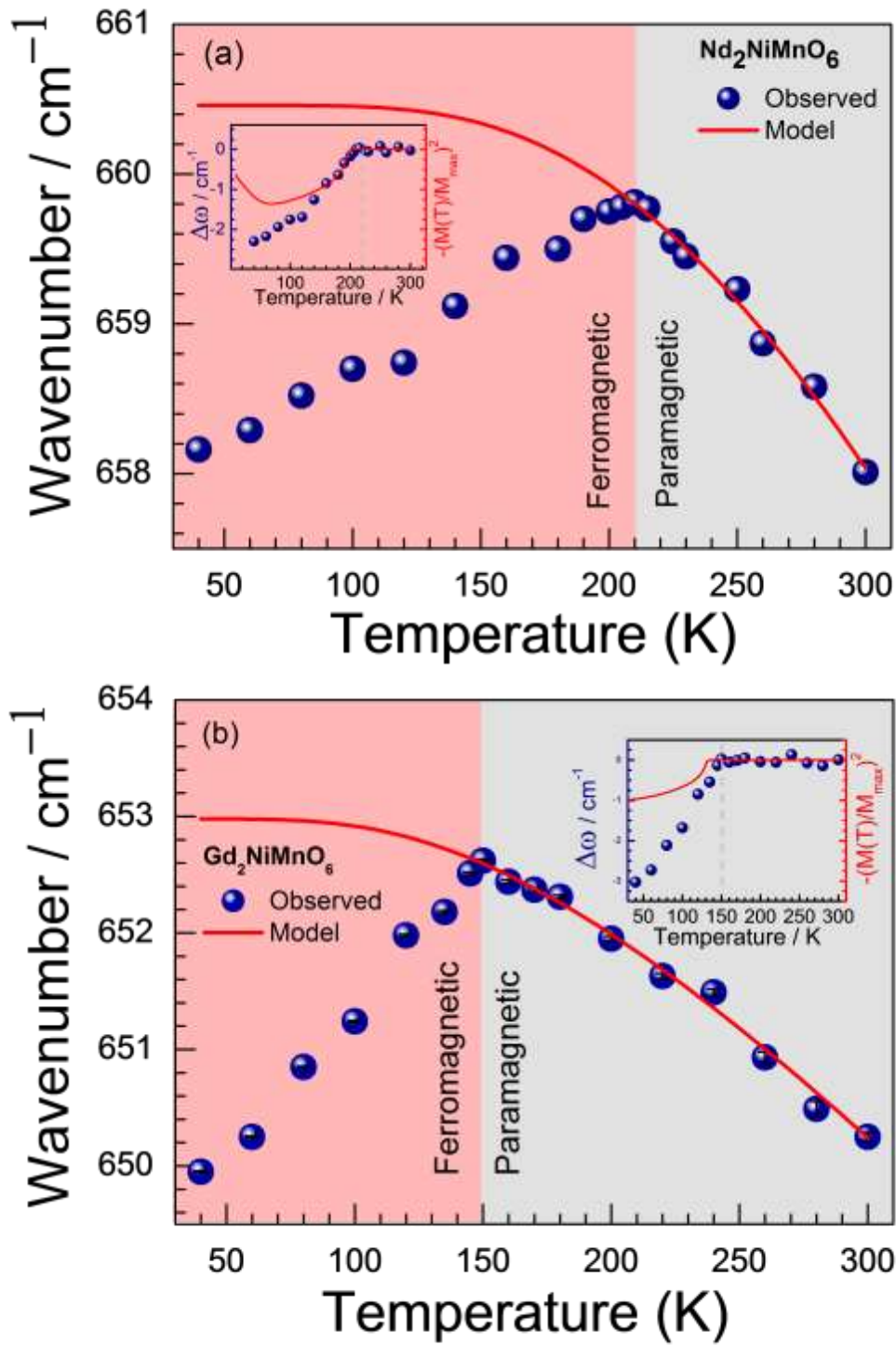

Figure 2 –Temperature dependence of the position of the stretching phonons in (a) NNMO and (b) GNMO. The insets in both figures indicate the temperature dependence of the deviation of the stretching phonon position in (a) NNMO and (b) GNMO with relation to $\left(\frac{M(T)}{M_o}\right)^2$ values obtained from magnetization measurements.

This effect is similar to the one observed in other ferromagnetic manganites, and it is associated to the renormalization of the phonons induced by the magnetic ordering [9,11–14,22] due to the coupling between magnetic ordering and lattice. According to Granado et al[23], in the absence magnetostriction effects and electronic states renormalization, the contribution of a spin-phonon coupling $(\Delta\omega_{ph})_k$ to the position change of the $k^{th}$ phonon is approximatelly given by

$$(\Delta\omega_{ph})_k \approx -\sum_{i,j>i} J_{ij}\langle \mathbf{S}_i \cdot \mathbf{S}_j\rangle \quad (1)$$

where $J_{ij}$ is the superexchange integral and $\langle \mathbf{S}_i \cdot \mathbf{S}_j\rangle$ is the spin correlation function. This contribution, considering only first neighboor interactions and a molecular field approximation, can be considered in the case of the stretching phonon, as

$$(\Delta\omega_{ph})_{stret.} \propto \left(\frac{M(T)}{M_o}\right)^2 \quad (2)$$

where $M(T)$ is the magnetization at the temperature T, and $M_o$ is the maximum magnetization. In the insets of the Figure 2 we showed the temperature dependence of the departure from the anharmonic behavior of the stretching bands for NNMO and GNMO compared with $\left(\frac{M(T)}{M_o}\right)^2$ obtained from Figure S2. We can see, at temperatures near the transitions the model agree very well with experimental data.

Comparing the spin-phonon coupling magnitude in RE$_2$NiMnO$_6$ ceramis for RE=Nd and Gd (this work), RE = Pr[11], Y[13] and Tb[12] as well as the B-site ordering in the respective compounds, the results suggest ionic radii does not influence significantly the spin-phonon coupling, with its magnitude being mainly controlled by the B-site ordering, as suggested by Macedo Filho et al[13]. Figure 3 shows a comparison considering temperature and wavenumber normalized for RE = Nd, Gd and Y [13]. Also, we showed the behavior for RE = La[xx] and Pr[yy] for sake of comparison. In this plot, the scales were renormalized following the changes $T \rightarrow T/T_C$ and $\omega \rightarrow \omega/\omega(T = 0)$. As we can see, for these all compounds the spin-phonon magnitude, here measured as $\Delta\omega_{max}$ in the normalized plot, is around 0,5% of $\omega_o$ for all compounds, being the differences within the resolution limit.

It is inportant point out that in the ferromagnetic phase the deviation of the anharmonic behavior follow a power law of the form

$$\Delta\omega \propto -\left(1 - \frac{T}{T_C}\right)^{\gamma}$$

as evidenced by dashed lines in Figure 3 ($T < T_C$). For all fitted compounds $1.3 < \gamma < 1.5$, suggesting a universal behavior independent of the rare earth or, in other words, of the unit cell size. The paramenter $\gamma$ is a normalized order when considering that the spin–spin correlation function $\langle S_i \cdot S_j \rangle$ can be described by mean field theory. This power law behavior is similar to that observed for orthocromites, as explicited by Amrani et al[??] for transverse optical phonons.

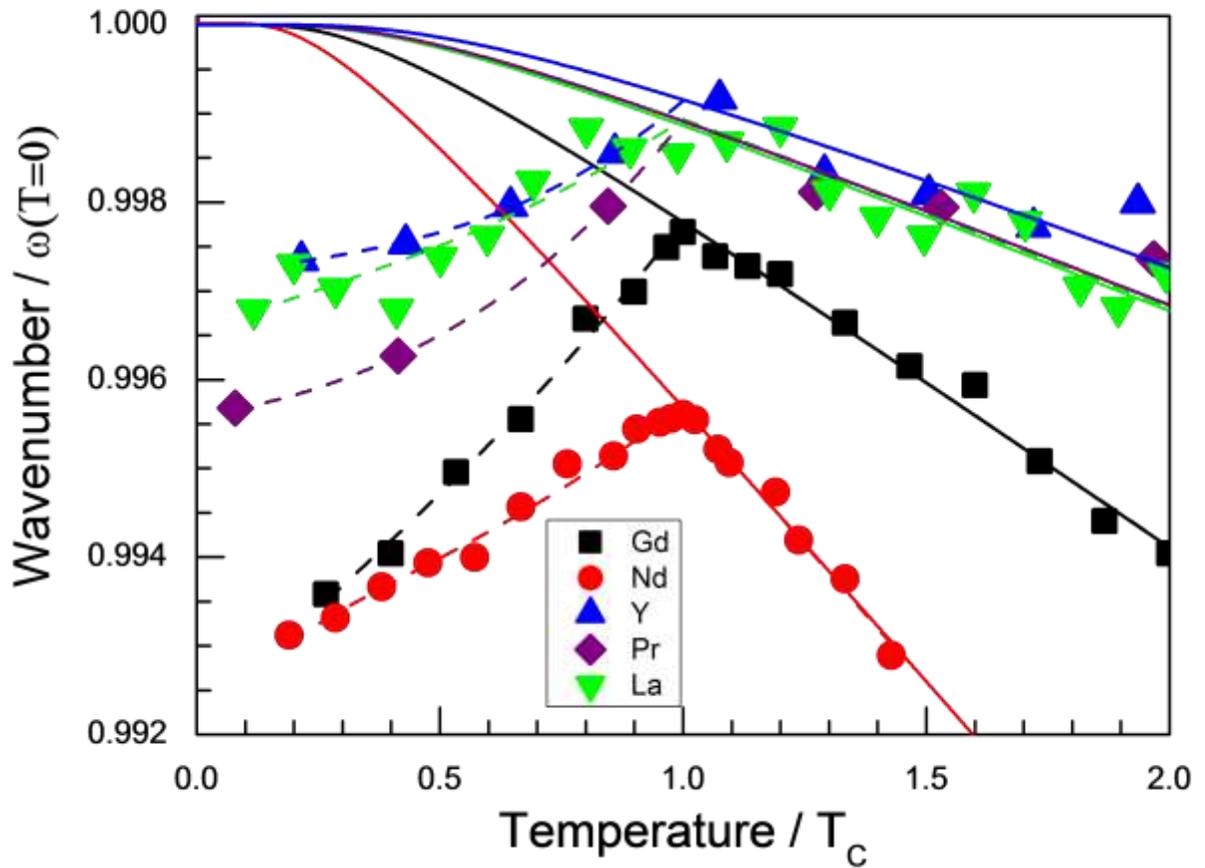

Figure 3 – Normalized plot comparing the spin-phonon magnitude for RE = Nd, Gd and Y[13]. Solid lines indicate the modeling according to the Balkanski's model. Dashed lines indicate a power law fitting of the phonon position deviation from the Balkanski's model.

In this letter, we showed rare earth manganites $RE_2NiMnO_6$ with double perovskite structure shows a spin-phonon coupling whose magnitude independent of the rare-earth ionic radius. Also, the deviation of the position of the stretching phonon in the ferromagnetic phase with relation to the anharmonic contributions follows a power law, whose order parameter is also independent of the rare earth ionic radius.

**Acknowledgements**

The authors acknowledge the financial support of the Brazilian funding agencies: CAPES, CNPq, FUNCAP, FAPEMA and IPDI (Instituto de Pesquisa, Desenvolvimento e Inovação, Ceará, Brazil).